\documentclass{article}
\usepackage[english]{babel}
\usepackage[a4paper,top=2cm,bottom=2cm,left=3cm,right=3cm,marginparwidth=1.75cm]{geometry}
\usepackage{graphicx}
\usepackage{natbib}

\usepackage{authblk}
\usepackage{appendix}
\usepackage{amsmath}
\usepackage{txfonts}
\usepackage{url}

\begin{document} 

\title{Photospheric index of geomagnetic activity}
\author[1,*]{Ismo Tähtinen}
\author[1]{Timo Qvick}
\author[1]{Timo Asikainen}
\author[1]{Kalevi Mursula}
\affil[1]{Space Physics and Astronomy Research Unit, University of Oulu, POB 8000, FI-90014, Oulu, Finland}
\affil[*]{Corresponding author: ismo.tahtinen@oulu.fi}

    \maketitle

  \abstract
   Geomagnetic indices quantify the disturbances of the Earth's magnetic field. They form an important space weather and space climate record that tracks solar-terrestrial effects on the Earth. Although geomagnetic activity has its origins in the photospheric magnetic fields, the two have so far not been directly related.
   We develop a photospheric index of geomagnetic activity (PIGA) that measures the geoeffectiveness of the Sun based on photospheric magnetic field as presented in synoptic magnetograms.
   We construct the PIGA from synoptic magnetograms of Wilcox Solar Observatory by relating the equatorial and axial dipole components of the photospheric magnetic field to the geomagnetic Kp index by means of linear regression.
   PIGA captures well the evolution of the geomagnetic activity over the past 50 years at the resolution of solar rotation, having linear correlation of 0.71 with the geomagnetic Kp index.
   PIGA makes it straightforward to measure the geoeffectiveness of the Sun using the dominant structure of the global solar magnetic field.
   PIGA can for example be used to study the geoeffectiveness of the Sun using historical simulations or the predicted solar magnetic field.
   PIGA also allows to use observed geomagnetic activity as an additional constraint for the early magnetic field reconstructions of the solar magnetic field.

\section{Introduction}
The solar magnetic field is responsible for a wide range of geoeffective phenomena that disturb the Earth's magnetic field.
These disturbances, driven by the heliospheric magnetic field and solar wind, are recorded in geomagnetic activity indices, some of which extend back to the 19th century \citep{Mayaud1971, Mayaud1972}.
Although geomagnetic activity ultimately arises from the solar magnetic field, a quantitative connection between the two has not yet been formulated.

Geomagnetic indices, which there are several, characterize the intensity and temporal evolution of geomagnetic activity  based on ground-level magnetic field measurements \citep{Rostoker1972, Mayaud1980}. 
Geomagnetic activity is driven by the interaction between the solar wind and the Earth's magnetic field which controls the entry of solar wind particles, energy, and magnetic flux into the Earth's magnetosphere.
At monthly and longer timescales this coupling of the solar wind and the magnetosphere depends on the heliospheric magnetic field strength $B$ and solar wind speed $V$.
For example, geomagnetic aa index correlates well with their product $BV$ or a more general power $BV^\alpha$ \citep[see, e.g.,][ and references therein]{Lockwood2022}.
$B$ and $V$ are typically intensified in solar wind structures related to coronal mass ejections (CMEs) and high-speed solar wind streams and their interaction regions, the two main drivers of geomagnetic activity \citep{LegrandSimon1989, Mursula2022, Richardson2012a}.

While CMEs are typically more localized phenomena on the Sun, related to the eruption of strong and complex magnetic fields of active regions, the occurrence of high-speed streams and the strength and structure of the heliospheric magnetic field are closely related to the global structure of the solar magnetic field.
High-speed streams predominantly originate from large unipolar regions called coronal holes, from which the magnetic field and solar wind can freely expand to the heliosphere \citep{Krieger1973, Richardson2018}.

The long-term evolution of geomagnetic activity differed from that of solar activity during the last century, which has been related to the changes in the global structure of the solar magnetic field \citep{Lockwood2022}.
Unlike solar activity, which started to decline in the 1960s, geomagnetic activity kept increasing or stayed at a relatively high level until the start of the new millennium \citep{Stamper1999,Lockwood1999}.
This may be related to an increase of low-latitude coronal holes towards the end of the century \citep{Gibson_2009,Mursula2017}.

The simplest approximation of the global solar magnetic field is a dipole, which corresponds to the first multipole of the spherical harmonic expansion of the radial photospheric magnetic field.
As the higher order multipoles rapidly fall-off with height, the solar dipole is especially important for the heliospheric magnetic field \citep{WangSheeley1988}.
Furthermore, because the solar dipole reflects the strength and location of large unipolar magnetic regions, it is also closely related to coronal holes and high-speed solar wind streams \citep{Hundhausen1977,Wang2009CH,Cranmer2009}.

The solar dipole can be decomposed to its equatorial and axial components.
The equatorial component fluctuates on a timescale of about one year, as it depends on continuous flux emergence.
It is also highly sensitive to the longitudinal distribution of active regions \citep{WangSheeley2003,Tahtinen2026a}.
On mid-term timescales, from a few solar rotations to a few years, the equatorial component has been found to play a major role in shaping the heliospheric magnetic field \citep{WangSheeley1988,WangSheeley2000a,WangSheeley2000b,WangSheeley2006}.
The axial component evolves much more steadily, varying in a rough antiphase with the sunspot cycle.
Around solar minima, when the solar dipole is almost entirely axial, persistent high-speed streams flow from the polar regions, although they typically do not reach Earth as effectively as the more variable low-latitude high-speed streams associated with the equatorial dipole \citep{Richardson2018}.

While the long-term geomagnetic activity has been qualitatively linked to the properties of the photospheric magnetic field, no quantitative connection has been formulated.
In this Letter, we present a photospheric index of geomagnetic activity (PIGA) that relates the dominant photospheric magnetic flux distribution to magnetic activity experienced at the Earth as described by the geomagnetic Kp index.
We construct this index from the equatorial and axial dipole components that can be straightforwardly calculated from synoptic magnetograms.

\section{Data}
We derive PIGA using the longest-running synoptic magnetogram series that has been measured at Wilcox Solar Observatory \citep[WSO;][]{Scherrer1977}.
The resolution of WSO data is 30 pixels in sine latitude and 72 pixels in longitude.
The data cover Carrington rotations 1642--2293 starting from the beginning of solar cycle 21 (May 1976) and reaching the recent solar maximum (January 2025).
We converted line-of-sight measurements into a pseudo-radial field by dividing with the cosine of latitude.

To represent the intensity and temporal evolution of geomagnetic activity, we use the Kp index, first introduced by \cite{Bartels1949}. 
The Kp index is one of the most important and widely-used geomagnetic indices, depicting the range of variation of the Earth's ground-level magnetic field compared to its quiet-time level at middle latitudes. 
It is quasi-logarithmic and it is expressed in thirds of a unit in the range of 0--9.
The Kp index is highly relevant especially for space climate studies since it is one of the most reliable and longest running indices, with values continuing from 1932 to present \citep{Matzka2021}.
We use daily values of the Kp index from the OMNI 2 database \citep{KingPapitashvili2005}, averaged over Carrington rotations (27.2753 days) to match the temporal resolution of the WSO data.

Additionally, we use magnetogram data from Mount Wilson Observatory \citep[MWO;][]{Howard1974}, from Kitt Peak \citep{Jones1992}, from the Michelson Doppler Imager on board the Solar and Heliospheric Observatory \citep[SOHO/MDI;][]{Scherrer1995}, from the Vector Spectromagnetograph on the Synoptic Optical Long-term Investigations of the Sun telescope \citep[SOLIS/VSM;][]{Keller2003}, and from the Helioseismic and Magnetic Imager on board the Solar Dynamics Observatory \citep[SDO/HMI;][]{Scherrer2012,Pesnell2012} to demonstrate the applicability of PIGA over different datasets.
 
\section{The photospheric index of geomagnetic activity}
\begin{figure}[!htbp]
\resizebox{\hsize}{!}{\includegraphics{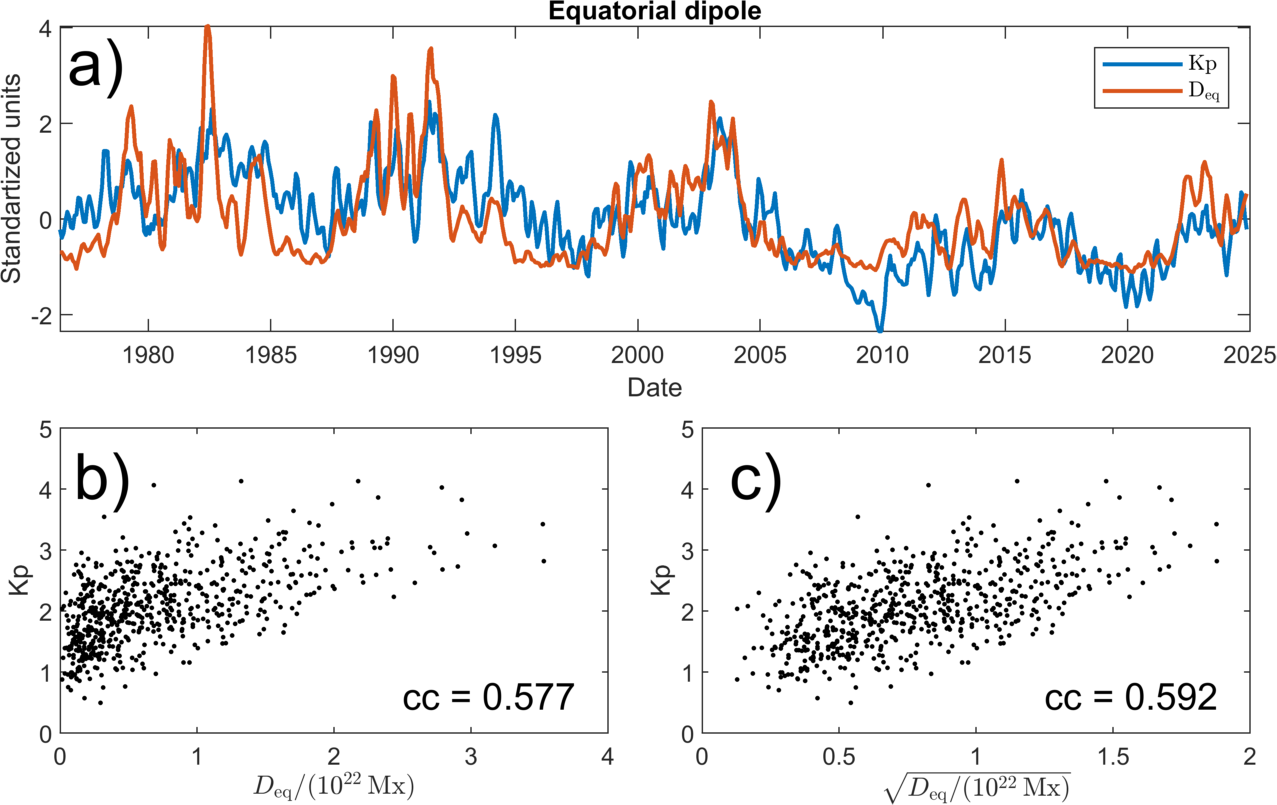}}
\caption{Kp index and the WSO equatorial dipole $D_{eq}$. 
a) Standardized time series smoothed with a 3-rotation moving average. 
b) Scatter plot of the Kp index versus $D_{eq}$. 
c) Scatter plot of the Kp index versus $\sqrt{D_{eq}}$. 
The dipole components are expressed in units of $10^{22}$~Mx.
Data in scatter plots has not been smoothed.
Linear correlation coefficient (cc) is shown in panels b and c.}
\label{fig:Deq}
\end{figure}

\begin{figure}[!htbp]
\resizebox{\hsize}{!}{\includegraphics{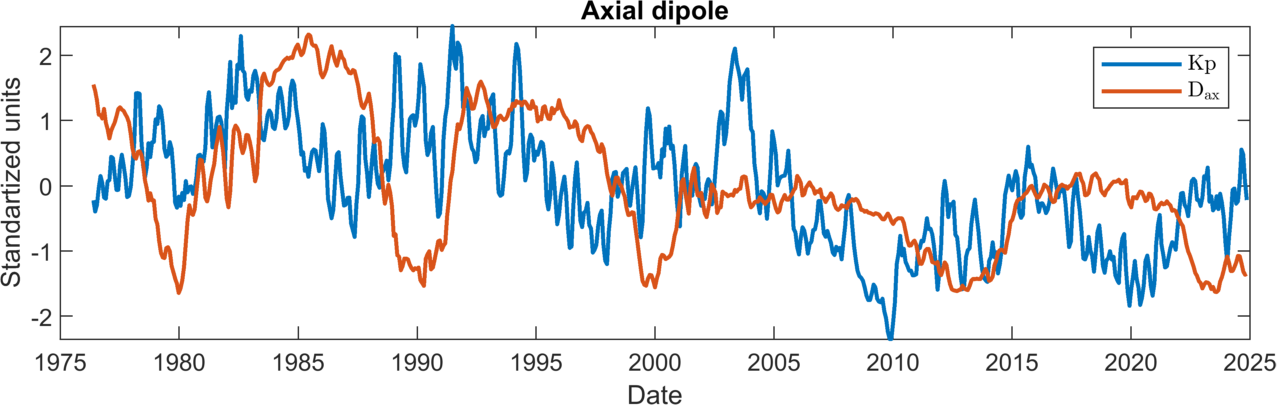}}
\caption{Kp index and the WSO axial dipole. The time series are standardized and smoothed with 3-rotation moving average.
}\label{fig:Dax}
\end{figure}

We construct PIGA from the equatorial, $D_{\rm eq}=|\textbf{D}|\cos{\lambda}$, and axial, $D_{\rm ax}=|\textbf{D}|\sin{|\lambda|}$, dipole components of the open dipole flux vector $\textbf{D}$, derived with the vector sum method \citep[][Appendix~\ref{appendix:VectorSum}]{Tahtinen2024,Tahtinen2026a}.
\textbf{D} has units of flux and equals spherical harmonic dipole up to a multiplicative factor  $\frac{4\pi R_\odot^2}{3}$.
Time series of the equatorial and axial dipole together with Kp index are shown in Figs.~\ref{fig:Deq} and \ref{fig:Dax}, respectively.
The dipole components are expressed in units of $10^{22}$~Mx. 

Figure~\ref{fig:Deq}a shows that much of geomagnetic variability at mid-term timescales from a few Carrington rotations to a few years is related to the fluctuations in the equatorial dipole.
However, the amplitude of these mid-term fluctuations in the equatorial component is more extreme than in the Kp index.
To reduce the amplitude of these spikes, we take the square root of the equatorial dipole component.
The two lower panels of Fig.~\ref{fig:Deq} show scatter plots of the Kp index as a function of $D_{\rm eq}$ (panel b) and $\sqrt{D_{\rm eq}}$ (panel c).
They clearly show that the relationship between Kp is more linear with $\sqrt{D_{\rm eq}}$.

Because the equatorial dipole essentially vanishes during each solar minimum, it does not capture the long-term declining trend in the baseline of geomagnetic activity. 
However, as this long-term trend is related to the weakening of solar polar fields \citep{Wang_2009}, it is present in the axial dipole component shown in Fig.~~\ref{fig:Dax}.
Thus we can construct PIGA as a linear combination of these two dipole components.

We define PIGA as
\begin{align}\label{eq:PIGA}
        \rm{PIGA} = \alpha_0+\alpha_1\sqrt{{\hat{D}_{\rm eq}}} + \alpha_2{\hat{D}_{\rm ax}},
\end{align}
where $\hat{D}_{\rm eq}$ and $\hat{D}_{\rm ax}$ are the dimensionless forms of
$D_{\rm eq}$ and $D_{\rm ax}$, obtained by expressing the dipole components in
units of $10^{22}\,\mathrm{Mx}$.
Coefficients $\alpha_0$, $\alpha_1$ and $\alpha_2$  are determined by linear regression against the geomagnetic Kp index.
Instead of ordinary least squares, we use the total least squares (TLS) fitting method that takes into account the errors/residual variability in all variables and in practice minimizes the norm of residuals orthogonal to the fit \citep[see, e.g.,][]{Golub1980}.

The regression coefficients for PIGA are given in Table~\ref{Table:RegressionCoefficients}, together with the linear correlation coefficient (cc) and mean absolute deviation (MAD).
The intercept term $\alpha_0=0.05$ is small and compatible with zero within one bootstrap standard error ($\sigma=0.07$).
The dipole coefficients, on the other hand, are clearly different from zero.
The equatorial term contributes about twice as much to PIGA as the axial term over the studied interval, as measured by the ratio $\sum(\alpha_1 \sqrt{\hat{D}_{\rm eq}})/\sum(\alpha_2 \hat{D}_{\rm ax})=2.1$.

\begin{table}
\caption{Coefficients of PIGA. 
The first three columns show the coefficients obtained with TLS regression, together with twice the standard error estimated from bootstrap sampling. 
The fourth column shows the linear correlation coefficient (cc), and the fifth column shows the mean absolute deviation (MAD).
}  
\centering
\begin{tabular}{c c c c c}
\hline\hline
$\alpha_0$ & $\alpha_1$ & $\alpha_2$ & cc & MAD\\
\hline                            
$ 0.05 \pm 0.14$ &  $1.8 \pm 0.1$  & $0.33 \pm 0.04 $ & 0.708 & 0.380 \\     
\hline                           
\end{tabular}               
\label{Table:RegressionCoefficients}
\end{table}

\begin{figure}[!htbp]
\resizebox{\hsize}{!}{\includegraphics{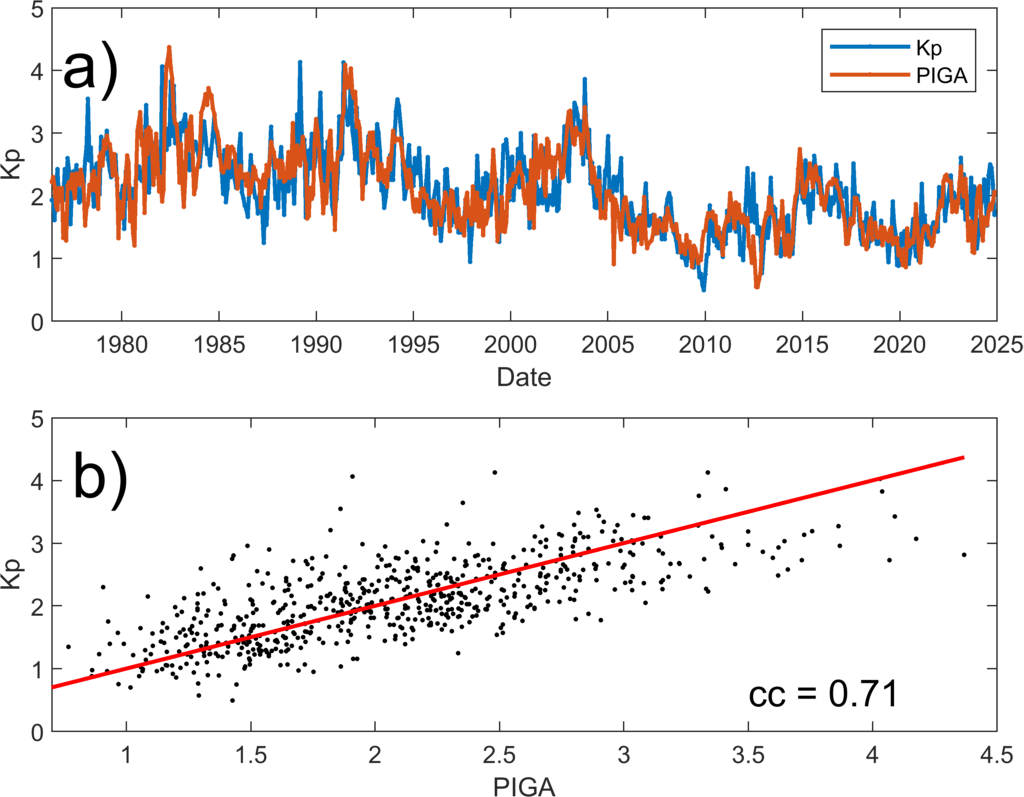}}
\caption{Comparison of PIGA and the Kp index. 
a) Time series of rotational values of PIGA (orange curve) and the Kp index (blue curve). 
b) Scatter plot of the Kp index versus PIGA. 
Red line shows the one-to-one relation, Kp=PIGA, for reference.
}\label{fig:PIGA}
\end{figure}

Figure~\ref{fig:PIGA} compares the Kp index and PIGA.
The time series show that PIGA captures the evolution of geomagnetic activity over the past 50 years quite well.
PIGA performs especially well with the long-term trend and during declining phases of cycle 22 on the early 1990s and cycle 24 on the late 2010s.
On the other hand, PIGA tends to over- or underestimate geomagnetic activity around solar maxima, for example near the turn of the millennium and in the early 2010s.

\begin{figure}[!htbp]
\resizebox{\hsize}{!}{\includegraphics{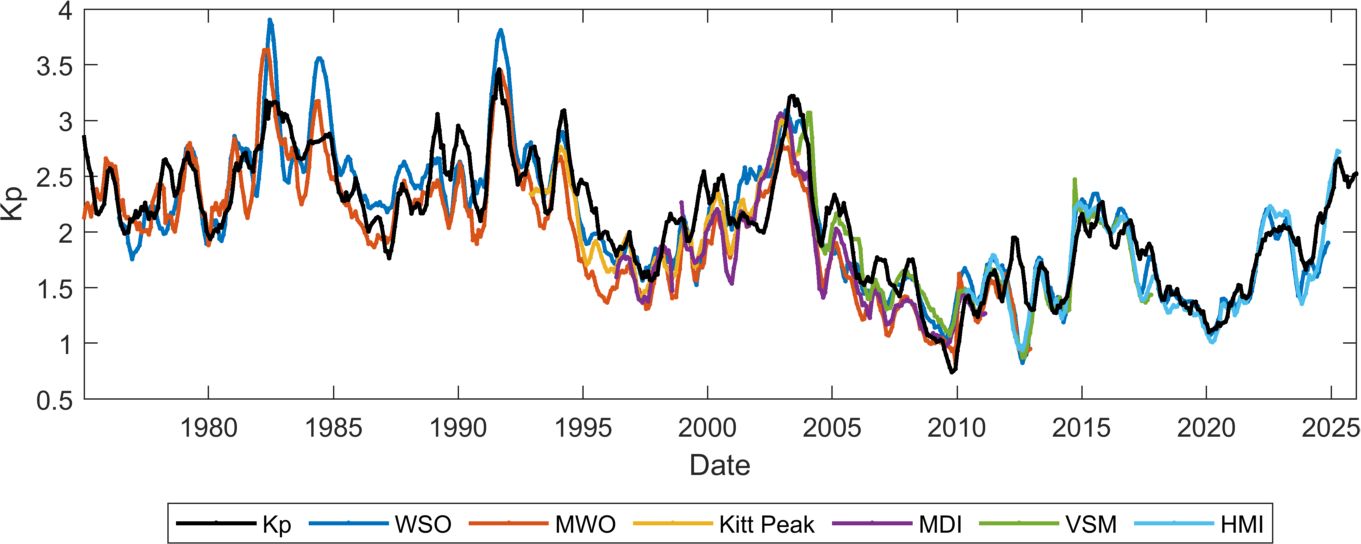}}
\caption{Kp index and PIGA from different magnetogram datasets. 
Data have been smoothed using a 7-rotation moving average.
}\label{fig:PIGA2}
\end{figure}

Regression coefficients derived here can also be used to calculate PIGA from other magnetogram datasets, provided that the dipole components are scaled to the level of WSO.
This can be done using the harmonic scaling coefficients of \citet{Virtanen2017}.
Figure~\ref{fig:PIGA2} shows that PIGA time series from several different magnetogram datasets agree well with each other.
Notably, PIGA from HMI magnetograms closely follows the Kp index in 2025, beyond the time frame of the WSO dataset.

\section{Discussion}
We have constructed PIGA, an index that relates the dominant large-scale magnetic flux distribution in photospheric synoptic maps to geomagnetic activity measured at the surface of Earth.
PIGA quantifies the geoeffectiveness of the photospheric magnetic field based on the equatorial and axial dipole components.
These dipole components reflect the geoeffectiveness in two ways.
First, because the heliospheric magnetic field is dominated by low-order multipoles \citep[][]{Wang2009CH}, their amplitudes reflect the strength of the heliospheric magnetic field.
Second, the relative strengths of the axial and equatorial components describe the inclination of the solar dipole, which is connected to the latitudinal organization of coronal holes and high-speed streams.
The equatorial dipole contributes about twice as much to PIGA as the axial component, which likely reflects the greater geoeffectiveness of low-latitude coronal holes \citep{Richardson2018}.

The most notable advantages of PIGA are that dipole components are straightforward to calculate and they are completely independent of the modeling of the coronal and heliospheric magnetic fields.
Also, because dipole components are robust measures of the large-scale magnetic field that scale well between different instruments \citep{Virtanen2017}, PIGA produces well-matching results across different magnetograms.
PIGA correlates reasonably well (cc=0.71) with geomagnetic activity as measured by the Kp index at timescales longer than a Carrington rotation. 
Therefore, PIGA is suitable for studying longer-term geomagnetic activity and space climate, rather than space weather at its typical timescales of a few days.

The simplicity of PIGA makes it very useful for estimating the geoeffectiveness of the solar magnetic field, for example, in surface flux transport (SFT) simulations \citep{Yeates2023}.
SFT simulations are often used to reconstruct historical solar magnetic fields from sunspot and calcium plage observations \citep{Cameron2010,Jiang2011,Virtanen2022,Yeates2025}.
They can also be used purely on statistical grounds, which is useful for predicting the future behavior of the solar magnetic field \citep{Cameron2016,Ijima2017,Jiang2018,Whitbread2018,Upton2018,Bhowmik2018}.
Because PIGA is constructed from the dipole components alone, a full-blown SFT simulation is not needed when calculating PIGA.
The dipole evolution can be estimated much more efficiently for example with the recent dipole flux transport (DFT) model \citep{Tahtinen2026b}.

One crucial problem with historical reconstructions based on direct observations of sunspots and calcium plages is the lack of independent constraints that can be used to measure their performance.
From the start of the 20th century, observations of polar faculae can be used to evaluate the performance of SFT simulations as their number is related to the strength of the polar fields \citep{MunozJaramillo2012,Yeates2025}.
The aa index of geomagnetic activity, which closely correlates with the Kp index used here, extends back to 1868 and can even be extended to 1844 using the Helsinki observatory data \citep{Nevanlinna_2004,Qvick2025}. With PIGA based on the aa index it will be possible to evaluate the performance of SFT simulations in these earlier times.
Furthermore, SFT models suffer from a well-known degeneracy in their parameter space, which polar field observations alone are not sufficient to alleviate \citep{Yeates2023}.
Recently, \citet{Tahtinen2026a} showed that the equatorial dipole component of the solar magnetic field can be used to uniquely constrain the SFT parameter space because it behaves oppositely with respect to diffusion compared to polar fields and other axisymmetric measures of the global magnetic field.
Because PIGA partly depends on the solar equatorial dipole, it opens up a possibility of using geomagnetic activity as an additional constraint for the parameter space of SFT models.
This offers an exciting possibility for providing better estimates of photospheric transport processes, such as diffusion, meridional flow, and differential rotation in the past.

PIGA can also be used to estimate some solar-terrestrial effects based on photospheric magnetic flux distributions.
This can be helpful for estimating the range of such effects that different topologies of the solar magnetic field can have.
This is useful for understanding both the past and the future of the near-Earth space environment.
Mid- to long-term geomagnetic activity, which is related to the global topology of the solar magnetic field, can affect the chemistry of the Earth's atmosphere by increasing energetic particle precipitation at high latitudes \citep{Funke2016, Salminen2020b}.
Due to this coupling to atmospheric dynamics, geomagnetic activity has been linked to wintertime temperatures \citep{Bochnicek_1998,Maliniemi2013,Seppala_2009,Thejll_2003}, as well as electricity consumption \citep{Juntunen2023} and production \citep{Juntunen2025a, Juntunen2026}. 
PIGA could be useful for seasonal prediction of wintertime weather, which only recently has started to take into account geomagnetic activity as a contributing factor \citep{Vokhmyanin2023, Vokhmyanin2025}.

\section{Conclusions}
Establishing a direct link between the global solar magnetic field and geomagnetic activity is crucial for understanding the range of geoeffectiveness of the solar magnetic field in the past and in the future.
PIGA offers a particularly simple method for estimating these effects at timescales longer than a Carrington rotation without the need for modeling the solar wind or the coronal or heliospheric magnetic fields.

\section*{Acknowledgements}
    The OMNI data were obtained from the GSFC/SPDF OMNIWeb interface at \url{https://omniweb.gsfc.nasa.gov}.
     Wilcox Solar Observatory data used in this study was obtained via the website \url{http://wso.stanford.edu} courtesy to J.T. Hoeksema.
    This study includes data from the synoptic program at the 150-foot Solar Tower of the Mt. Wilson Observatory.
The Mt. Wilson 150-Foot Solar Tower is operated by UCLA, with funding from NASA, ONR and NSF, under agreement with the Mt. Wilson Institute.
NSO/Kitt Peak magnetic data used here are produced cooperatively by NSF/NOAO, NASA/GSFC and NOAA/SEL.
Data were acquired by SOLIS instruments operated by NISP/NSO/AURA/NSF. SOHO/MDI is a project of international cooperation between ESA and NASA.
HMI data are courtesy of the Joint Science Operations Center (JSOC) Science Data Processing team at Stanford University.

\bibliography{bibliography} 

@string{AG = {Ann.\ Geophys.}}

@string{JASTP = {J. Atm.\ Sol.-Terr.\ Phys.}}

@string{JGRA = {J.\ Geophys.\ Res. (Atmosphere)}}

@string{JGRS = {J.\ Geophys.\ Res. (Space)}}

@incollection{Bartels1949,
	author = {Bartels, J.},
	title = {{Appendix B: The standardized index, Ks, and the planetary index, Kp}},
	booktitle = {IATME Bull. 12b},
	pages = {97--120},
	year = {1949},
    publisher = {{IUGG}},
}

@article{Bochnicek_1998,
  author  = {Bochn{\'\i}{\v c}ek, Josef and Bucha, Vladim{\'\i}r and Hejda, Pavel and Pycha, Jan},
  title   = {Relation Between Northern Hemisphere Winter Temperatures and Geomagnetic or Solar Activity at Different QBO Phases},
  journal = JASTP,
  volume  = {60},
  number  = {2},
  pages   = {219--232},
  year    = {1998},
  doi     = {10.1016/S1364-6826(97)00109-X}
}

@article{Funke2016,
	author = {Funke, B. and L\'opez-Puertas, M. and Stiller, G. P. and Versick, S. and von Clarmann, T.},
	title = {{A semi-empirical model for mesospheric and stratospheric NO$_{y}$ produced by energetic particle precipitation}},
	journal = {Atmospheric Chemistry and Physics},
	volume = {16},
	year = {2016},
	number = {13},
	pages = {8667--8693},
	url = {https://acp.copernicus.org/articles/16/8667/2016/},
	doi = {10.5194/acp-16-8667-2016}
}

@article{Gibson_2009,
   author = {{Gibson}, S.~E. and {Kozyra}, J.~U. and {de Toma}, G. and {Emery}, B.~A. and 
	{Onsager}, T. and {Thompson}, B.~J.},
    title = "{If the Sun is so quiet, why is the Earth ringing? A comparison of two solar minimum intervals}",
  journal = JGRS,
  year = 2009,
  volume = 114,
   number = {A13},
    pages = {A09105},
      doi = {10.1029/2009JA014342}
}

@ARTICLE{Golub1980,
       author = {{Golub}, Gene H. and {van Loan}, Charles F.},
        title = "{An Analysis of the Total Least Squares Problem}",
      journal = {SIAM Journal on Numerical Analysis},
         year = 1980,
        month = dec,
       volume = {17},
       number = {6},
        pages = {883-893},
          doi = {10.1137/0717073},
       adsurl = {https://ui.adsabs.harvard.edu/abs/1980SJNA...17..883G}
}

@article{Juntunen2023,
    author = {Juntunen, Veera and Asikainen, Timo},
    title = {{Electricity consumption in Finland influenced by climate effects of energetic particle precipitation}},
    journal = {Scientific Reports},
    year = {2023},
    OPTmonth = {Nov},
    OPTday = {23},
    volume = {13},
    number = {1},
    pages = {20546},
    issn = {2045-2322},
    doi = {10.1038/s41598-023-47605-8},
    url = {https://doi.org/10.1038/s41598-023-47605-8}
}

@article{Juntunen2025a,
	author = {Juntunen, V. and Asikainen, T. and Salminen, A.},
	title = {Influence of Energetic Electron Precipitation on Wind Power Generation in European Countries Mediated by the Polar Vortex},
	journal = {Space Weather},
	volume = {23},
	number = {5},
	pages = {e2024SW004157},
	doi = {https://doi.org/10.1029/2024SW004157},
	url = {https://agupubs.onlinelibrary.wiley.com/doi/abs/10.1029/2024SW004157},
	eprint = {https://agupubs.onlinelibrary.wiley.com/doi/pdf/10.1029/2024SW004157},
	OPTnote = {e2024SW004157 2024SW004157},
	year = {2025}
}

@article{Juntunen2026,
	author = {Juntunen, Veera and Asikainen, Timo and Salminen, Antti},
	title = {{Impact of the northern polar vortex and geomagnetic activity on solar radiation at surface and solar power production in Europe}},
	journal = {Environmental Research Letters},
	year = {2026},
	OPTmonth = {apr},
	publisher = {IOP Publishing},
	volume = {21},
	number = {8},
	pages = {084013},
	doi = {10.1088/1748-9326/ae5faf},
	url = {https://doi.org/10.1088/1748-9326/ae5faf}
}

@article{KingPapitashvili2005,
	author = {King, J. H. and Papitashvili, N. E.},
	title = {Solar wind spatial scales in and comparisons of hourly Wind and ACE plasma and magnetic field data},
	journal = {Journal of Geophysical Research: Space Physics},
    year = {2005},
    month = feb,
    volume = {110},
    number = {A2},
    eid = {A02104},
    pages = {A02104},
    doi = {10.1029/2004JA010649},
	url = {https://agupubs.onlinelibrary.wiley.com/doi/abs/10.1029/2004JA010649},
	eprint = {https://agupubs.onlinelibrary.wiley.com/doi/pdf/10.1029/2004JA010649}
}

@article{Krieger1973,
       author = {{Krieger}, A.~S. and {Timothy}, A.~F. and {Roelof}, E.~C.},
        title = {A Coronal Hole and Its Identification as the Source of a High Velocity Solar Wind Stream},
      journal = {Solar Physics},
         year = {1973},
     OPTmonth = {apr},
       volume = {29},
        issue = {2},
        pages = {505-525},
          doi = {10.1007/BF00150828},
       adsurl = {https://ui.adsabs.harvard.edu/abs/1973SoPh...29..505K}
}

@article{LegrandSimon1989,
       author = {Legrand, J. -P. and Simon, P.~A.},
        title = "{Solar cycle and geomagnetic activity: A review for geophysicists. Part 1. The contributions to geomagnetic activity of shock waves and of the solar wind.}",
      journal = {Annales Geophysicae},
         year = {1989},
     OPTmonth = {dec},
       volume = {7},
        pages = {565-593},
       adsurl = {https://ui.adsabs.harvard.edu/abs/1989AnGeo...7..565L}
}

@ARTICLE{Lockwood1999,
       author = {{Lockwood}, M. and {Stamper}, R. and {Wild}, M.~N.},
        title = "{A doubling of the Sun's coronal magnetic field during the past 100 years}",
      journal = {\nat},
         year = 1999,
        month = jun,
       volume = {399},
       number = {6735},
        pages = {437-439},
          doi = {10.1038/20867},
       adsurl = {https://ui.adsabs.harvard.edu/abs/1999Natur.399..437L}
}

@ARTICLE{Lockwood2022,
       author = {{Lockwood}, Mike and {Owens}, Mathew J. and {Barnard}, Luke A. and {Scott}, Chris J. and {Frost}, Anna M. and {Yu}, Bingkun and {Chi}, Yutian},
        title = "{Application of historic datasets to understanding open solar flux and the 20th-century grand solar maximum. 1. Geomagnetic, ionospheric, and sunspot observations}",
      journal = {Frontiers in Astronomy and Space Sciences},
         year = 2022,
        month = sep,
       volume = {9},
          eid = {960775},
        pages = {960775},
          doi = {10.3389/fspas.2022.960775},
       adsurl = {https://ui.adsabs.harvard.edu/abs/2022FrASS...9.0775L}
}

@article{Maliniemi2013,
	author = {Maliniemi, V. and Asikainen, T. and Mursula, K. and Seppälä, A.},
	title = {{QBO-dependent relation between electron precipitation and wintertime surface temperature}},
	journal = {Journal of Geophysical Research: Atmospheres},
	volume = {118},
	number = {12},
	pages = {6302-6310},
	doi = {https://doi.org/10.1002/jgrd.50518},
	url = {https://agupubs.onlinelibrary.wiley.com/doi/abs/10.1002/jgrd.50518},
	eprint = {https://agupubs.onlinelibrary.wiley.com/doi/pdf/10.1002/jgrd.50518},
	year = {2013}
}

@article{Matzka2021,
	author = {Matzka, J. and Stolle, C. and Yamazaki, Y. and Bronkalla, O. and Morschhauser, A.},
	title = {{The geomagnetic Kp index and derived indices of geomagnetic activity}},
	journal = {Space Weather},
	volume = {19},
	number = {5},
	pages = {e2020SW002641},
	doi = {https://doi.org/10.1029/2020SW002641},
	url = {https://agupubs.onlinelibrary.wiley.com/doi/abs/10.1029/2020SW002641},
	eprint = {https://agupubs.onlinelibrary.wiley.com/doi/pdf/10.1029/2020SW002641},
	OPTnote = {e2020SW002641 2020SW002641},
	year = {2021}
}

@article{Mayaud1971,
    author = {Mayaud, Pierre-Noël},
    title = {{Une mesure planétaire d'activité magnétique basée sur deux observatoires antipodaux}},
    journal = {Annales de Géophysique},
    volume = {27},
    year = {1971},
    number = {1},
    pages = {67--70},
    url = {},
    doi = {}
}

@article{Mayaud1972,
	author = {Mayaud, Pierre-No{\"e}l},
	title = {{The aa indices: A 100-year series characterizing the magnetic activity}},
	journal = {Journal of Geophysical Research (1896-1977)},
	volume = {77},
	number = {34},
	pages = {6870-6874},
	doi = {10.1029/JA077i034p06870},
	url = {https://agupubs.onlinelibrary.wiley.com/doi/abs/10.1029/JA077i034p06870},
	eprint = {https://agupubs.onlinelibrary.wiley.com/doi/pdf/10.1029/JA077i034p06870},
	year = {1972}
}

@book{Mayaud1980,
	title = {{Derivation, Meaning, and Use of Geomagnetic Indices}},
	author = {Mayaud, Pierre-No{\"e}l},
	isbn = {9781118663837},
	publisher = {American Geophysical Union (AGU)},
	series = {Geophys. Monogr. Ser.},
	volume = {22},
	doi = {10.1029/GM022},
	url = {https://agupubs.onlinelibrary.wiley.com/doi/book/10.1029/GM022},
	year = {1980},
}

@ARTICLE{MunozJaramillo2012,
       author = {{Mu{\~n}oz-Jaramillo}, Andr{\'e}s and {Sheeley}, Neil R. and {Zhang}, Jie and {DeLuca}, Edward E.},
        title = "{Calibrating 100 Years of Polar Faculae Measurements: Implications for the Evolution of the Heliospheric Magnetic Field}",
      journal = {\apj},
         year = 2012,
        month = jul,
       volume = {753},
       number = {2},
          eid = {146},
        pages = {146},
          doi = {10.1088/0004-637X/753/2/146},
archivePrefix = {arXiv},
       eprint = {1303.0345},
 primaryClass = {astro-ph.SR},
       adsurl = {https://ui.adsabs.harvard.edu/abs/2012ApJ...753..146M}
}

@ARTICLE{Mursula2017,
       author = {{Mursula}, K. and {Holappa}, L. and {Lukianova}, R.},
        title = "{Seasonal solar wind speeds for the last 100 years: Unique coronal hole structures during the peak and demise of the Grand Modern Maximum}",
      journal = {\grl},
         year = 2017,
        month = jan,
       volume = {44},
       number = {1},
        pages = {30-36},
          doi = {10.1002/2016GL071573},
archivePrefix = {arXiv},
       eprint = {1612.04941},
 primaryClass = {physics.space-ph},
       adsurl = {https://ui.adsabs.harvard.edu/abs/2017GeoRL..44...30M}
}

@article{Mursula2022,
    author = {Mursula, Kalevi and Qvick, Timo and Holappa, Lauri and Asikainen, Timo},
    title = {Magnetic Storms During the Space Age: Occurrence and Relation to Varying Solar Activity},
    journal = {Journal of Geophysical Research: Space Physics},
    volume = {127},
    number = {12},
    pages = {e2022JA030830},
    doi = {https://doi.org/10.1029/2022JA030830},
    url = {https://agupubs.onlinelibrary.wiley.com/doi/abs/10.1029/2022JA030830},
    eprint = {https://agupubs.onlinelibrary.wiley.com/doi/pdf/10.1029/2022JA030830},
    OPTnote = {e2022JA030830 2022JA030830},
    year = {2022}
}

@article{Nevanlinna_2004,
  author    = {Heikki Nevanlinna},
  title     = {Results of the Helsinki magnetic observatory 1844--1912},
  journal   = AG,
  year      = {2004},
  volume    = {22},
  number    = {5},
  pages     = {1691--1704},
  doi       = {10.5194/angeo-22-1691-2004},
  url       = {https://angeo.copernicus.org/articles/22/1691/2004/}
}

@article{Qvick2025,
    author = {Qvick, Timo and Asikainen, Timo and Mursula, Kalevi},
    title = {Predicting Geomagnetic Activity Cycles},
    journal = {Space Weather},
    volume = {23},
    number = {5},
    pages = {e2024SW004074},
    doi = {https://doi.org/10.1029/2024SW004074},
    OPTurl = {https://agupubs.onlinelibrary.wiley.com/doi/abs/10.1029/2024SW004074},
    eprint = {https://agupubs.onlinelibrary.wiley.com/doi/pdf/10.1029/2024SW004074},
    OPTnote = {e2024SW004074 2024SW004074},
    year = {2025}
}

@article{Richardson2012a,
	author = {Richardson, Ian G. and Cane, Hilary V.},
	title = {Solar wind drivers of geomagnetic storms during more than four solar cycles},
	doi = "10.1051/swsc/2012001",
	url = "http://dx.doi.org/10.1051/swsc/2012001",
	journal = {J. Space Weather Space Clim.},
	year = 2012,
	volume = 2,
	pages = "A01",
}

@article{Richardson2018,
	author = {Richardson, Ian G.},
	title = {Solar wind stream interaction regions throughout the heliosphere},
	journal = {Living Reviews in Solar Physics},
	year = {2018},
	volume = {15},
	number = {1},
	pages = {1},
	issn = {1614-4961},
	doi = {10.1007/s41116-017-0011-z},
	url = {https://doi.org/10.1007/s41116-017-0011-z}
}

@article{Rostoker1972,
    author = {Rostoker, Gordon},
    title = {Geomagnetic indices},
    journal = {Reviews of Geophysics},
    volume = {10},
    number = {4},
    pages = {935-950},
    doi = {https://doi.org/10.1029/RG010i004p00935},
    url = {https://agupubs.onlinelibrary.wiley.com/doi/abs/10.1029/RG010i004p00935},
    eprint = {https://agupubs.onlinelibrary.wiley.com/doi/pdf/10.1029/RG010i004p00935},
    year = {1972}
}

@article{Salminen2020b,
	author = {{Salminen}, Antti and {Asikainen}, Timo and {Maliniemi}, Ville and {Mursula}, Kalevi},
	title = {Comparing the effects of solar-related and terrestrial drivers on the northern polar vortex},
	journal = {J. Space Weather Space Clim.},
	year = 2020,
	volume = 10,
	pages = {56},
	doi = {10.1051/swsc/2020058},
	url = {https://doi.org/10.1051/swsc/2020058},
}

@ARTICLE{Scherrer1977,
       author = {{Scherrer}, P.~H. and {Wilcox}, J.~M. and {Svalgaard}, L. and {Duvall}, T.~L., Jr. and {Dittmer}, P.~H. and {Gustafson}, E.~K.},
        title = "{The mean magnetic field of the Sun: observations at Stanford.}",
      journal = {\solphys},
         year = 1977,
        month = oct,
       volume = {54},
       number = {2},
        pages = {353-361},
          doi = {10.1007/BF00159925},
       adsurl = {https://ui.adsabs.harvard.edu/abs/1977SoPh...54..353S}
}

@article{Seppala_2009,
   author = {A. Seppälä and C.E. Randall and M.A. Clilverd and E. Rozanov and C.J. Rodger},
    title = "{Geomagnetic activity and polar surface air temperature variability}",
journal = JGRS,
     year = 2009,
   volume = 114,
    pages = "A10312,doi:10.1029/2008JA014029"
}

@ARTICLE{Stamper1999,
       author = {{Stamper}, R. and {Lockwood}, M. and {Wild}, M.~N. and {Clark}, T.~D.~G.},
        title = "{Solar causes of the long-term increase in geomagnetic activity}",
      journal = {\jgr},
         year = 1999,
        month = jan,
       volume = {104},
       number = {A12},
        pages = {28325-28342},
          doi = {10.1029/1999JA900311},
       adsurl = {https://ui.adsabs.harvard.edu/abs/1999JGR...10428325S}
}

@ARTICLE{Tahtinen2024,
       author = {{T{\"a}htinen}, Ismo and {Asikainen}, Timo and {Mursula}, Kalevi},
        title = "{Straight outta photosphere: Open solar flux without coronal modeling}",
      journal = {\aap},
         year = 2024,
        month = aug,
       volume = {688},
          eid = {L32},
        pages = {L32},
          doi = {10.1051/0004-6361/202451267},
archivePrefix = {arXiv},
       eprint = {2408.11525},
 primaryClass = {astro-ph.SR},
       adsurl = {https://ui.adsabs.harvard.edu/abs/2024A&A...688L..32T}
}

@ARTICLE{Tahtinen2026a,
       author = {{T{\"a}htinen}, Ismo and {Asikainen}, Timo and {Mursula}, Kalevi},
        title = "{Active regions and the large-scale magnetic field of solar cycle 24}",
      journal = {\aap},
         year = 2026,
        month = feb,
       volume = {706},
          eid = {A235},
        pages = {A235},
          doi = {10.1051/0004-6361/202557466},
       adsurl = {https://ui.adsabs.harvard.edu/abs/2026A&A...706A.235T}
}

@ARTICLE{Tahtinen2026b,
       author = {{T{\"a}htinen}, Ismo and {Asikainen}, Timo and {Mursula}, Kalevi},
        title = "{Ultra-fast simulations of the solar dipole and open flux}",
      journal = {\aap},
         year = 2026,
        month = apr,
       volume = {708},
          eid = {L21},
        pages = {L21},
          doi = {10.1051/0004-6361/202659586},
       adsurl = {https://ui.adsabs.harvard.edu/abs/2026A&A...708L..21T}
}

@article{Thejll_2003,
  author  = {Thejll, Peter and Cliver, Edward W. and Gleisner, Hans},
  title   = {On the Relation Between Geomagnetic Activity and Lower Tropospheric Climate},
  journal = JGRA,
  volume  = {108},
  number  = {D18},
  pages   = {4589},
  year    = {2003},
  doi     = {10.1029/2003JD003562}
}

@article{Vokhmyanin2023,
    author = {Vokhmyanin, Mikhail and Asikainen, Timo and Salminen, Antti and Mursula, Kalevi},
    title = {Long-Term Prediction of Sudden Stratospheric Warmings With Geomagnetic and Solar Activity},
    journal = {Journal of Geophysical Research: Atmospheres},
    volume = {128},
    number = {6},
    pages = {e2022JD037337},
    doi = {https://doi.org/10.1029/2022JD037337},
    url = {https://agupubs.onlinelibrary.wiley.com/doi/abs/10.1029/2022JD037337},
    eprint = {https://agupubs.onlinelibrary.wiley.com/doi/pdf/10.1029/2022JD037337},
    OPTnote = {e2022JD037337 2022JD037337},
    year = {2023}
}

@article{Vokhmyanin2025,
	author = {Vokhmyanin, Mikhail and Asikainen, Timo and Salminen, Antti and Mursula, Kalevi},
	title = {Temperature Anomalies During Late Boreal Winters With and Without Sudden Stratospheric Warming},
	journal = {Geophysical Research Letters},
	volume = {52},
	number = {2},
	pages = {e2024GL110803},
	doi = {https://doi.org/10.1029/2024GL110803},
	url = {https://agupubs.onlinelibrary.wiley.com/doi/abs/10.1029/2024GL110803},
	eprint = {https://agupubs.onlinelibrary.wiley.com/doi/pdf/10.1029/2024GL110803},
	OPTnote = {e2024GL110803 2024GL110803},
	year = {2025}
}

@ARTICLE{Cameron2010,
       author = {{Cameron}, R.~H. and {Jiang}, J. and {Schmitt}, D. and {Sch{\"u}ssler}, M.},
        title = "{Surface Flux Transport Modeling for Solar Cycles 15-21: Effects of Cycle-Dependent Tilt Angles of Sunspot Groups}",
      journal = {\apj},
         year = 2010,
        month = aug,
       volume = {719},
       number = {1},
        pages = {264-270},
          doi = {10.1088/0004-637X/719/1/264},
archivePrefix = {arXiv},
       eprint = {1006.3061},
 primaryClass = {astro-ph.SR},
       adsurl = {https://ui.adsabs.harvard.edu/abs/2010ApJ...719..264C}
}

@ARTICLE{Cameron2016,
       author = {{Cameron}, R.~H. and {Jiang}, J. and {Sch{\"u}ssler}, M.},
        title = "{Solar Cycle 25: Another Moderate Cycle?}",
      journal = {\apjl},
         year = 2016,
        month = jun,
       volume = {823},
       number = {2},
          eid = {L22},
        pages = {L22},
          doi = {10.3847/2041-8205/823/2/L22},
archivePrefix = {arXiv},
       eprint = {1604.05405},
 primaryClass = {astro-ph.SR},
       adsurl = {https://ui.adsabs.harvard.edu/abs/2016ApJ...823L..22C}
}

@ARTICLE{Jiang2011,
       author = {{Jiang}, J. and {Cameron}, R.~H. and {Schmitt}, D. and {Sch{\"u}ssler}, M.},
        title = "{The solar magnetic field since 1700. II. Physical reconstruction of total, polar and open flux}",
      journal = {\aap},
         year = 2011,
        month = apr,
       volume = {528},
          eid = {A83},
        pages = {A83},
          doi = {10.1051/0004-6361/201016168},
archivePrefix = {arXiv},
       eprint = {1102.1270},
 primaryClass = {astro-ph.SR},
       adsurl = {https://ui.adsabs.harvard.edu/abs/2011A&A...528A..83J}
}

@ARTICLE{Ijima2017,
       author = {{Iijima}, H. and {Hotta}, H. and {Imada}, S. and {Kusano}, K. and {Shiota}, D.},
        title = "{Improvement of solar-cycle prediction: Plateau of solar axial dipole moment}",
      journal = {\aap},
         year = 2017,
        month = nov,
       volume = {607},
          eid = {L2},
        pages = {L2},
          doi = {10.1051/0004-6361/201731813},
archivePrefix = {arXiv},
       eprint = {1710.06528},
 primaryClass = {astro-ph.SR},
       adsurl = {https://ui.adsabs.harvard.edu/abs/2017A&A...607L...2I}
}

@ARTICLE{Jiang2018,
       author = {{Jiang}, Jie and {Wang}, Jing-Xiu and {Jiao}, Qi-Rong and {Cao}, Jin-Bin},
        title = "{Predictability of the Solar Cycle Over One Cycle}",
      journal = {\apj},
         year = 2018,
        month = aug,
       volume = {863},
       number = {2},
          eid = {159},
        pages = {159},
          doi = {10.3847/1538-4357/aad197},
archivePrefix = {arXiv},
       eprint = {1807.01543},
 primaryClass = {astro-ph.SR},
       adsurl = {https://ui.adsabs.harvard.edu/abs/2018ApJ...863..159J}
}

@ARTICLE{Virtanen2022,
       author = {{Virtanen}, I.~O.~I. and {Pevtsov}, A.~A. and {Bertello}, L. and {Mursula}, K.},
        title = "{Reconstructing solar magnetic fields from historical observations. IX. The photospheric magnetic field from 1915 to 1985}",
      journal = {\aap},
         year = 2022,
        month = nov,
       volume = {667},
          eid = {A168},
        pages = {A168},
          doi = {10.1051/0004-6361/202244372},
       adsurl = {https://ui.adsabs.harvard.edu/abs/2022A&A...667A.168V}
}

@ARTICLE{Upton2018,
       author = {{Upton}, Lisa A. and {Hathaway}, David H.},
        title = "{An Updated Solar Cycle 25 Prediction With AFT: The Modern Minimum}",
      journal = {\grl},
         year = 2018,
        month = aug,
       volume = {45},
       number = {16},
        pages = {8091-8095},
          doi = {10.1029/2018GL078387},
archivePrefix = {arXiv},
       eprint = {1808.04868},
 primaryClass = {astro-ph.SR},
       adsurl = {https://ui.adsabs.harvard.edu/abs/2018GeoRL..45.8091U}
}

@ARTICLE{Bhowmik2018,
       author = {{Bhowmik}, Prantika and {Nandy}, Dibyendu},
        title = "{Prediction of the strength and timing of sunspot cycle 25 reveal decadal-scale space environmental conditions}",
      journal = {Nature Communications},
         year = 2018,
        month = dec,
       volume = {9},
          eid = {5209},
        pages = {5209},
          doi = {10.1038/s41467-018-07690-0},
archivePrefix = {arXiv},
       eprint = {1909.04537},
 primaryClass = {astro-ph.SR},
       adsurl = {https://ui.adsabs.harvard.edu/abs/2018NatCo...9.5209B}
}

@ARTICLE{WangSheeley2000a,
       author = {{Wang}, Y. -M. and {Lean}, J. and {Sheeley}, Jr., N.~R.},
        title = "{The long-term variation of the Sun's open magnetic flux}",
      journal = {\grl},
         year = 2000,
        month = feb,
       volume = {27},
       number = {4},
        pages = {505-508},
          doi = {10.1029/1999GL010744},
       adsurl = {https://ui.adsabs.harvard.edu/abs/2000GeoRL..27..505W}
}

@ARTICLE{WangSheeley2000b,
       author = {{Wang}, Y. -M. and {Sheeley}, Jr., N.~R. and {Lean}, J.},
        title = "{Understanding the evolution of the Sun's open magnetic flux}",
      journal = {\grl},
         year = 2000,
        month = mar,
       volume = {27},
       number = {5},
        pages = {621-624},
          doi = {10.1029/1999GL010759},
       adsurl = {https://ui.adsabs.harvard.edu/abs/2000GeoRL..27..621W}
}

@ARTICLE{WangSheeley2003,
       author = {{Wang}, Y. -M. and {Sheeley}, Jr., N.~R.},
        title = "{On the Fluctuating Component of the Sun's Large-Scale Magnetic Field}",
      journal = {\apj},
         year = 2003,
        month = jun,
       volume = {590},
       number = {2},
        pages = {1111-1120},
          doi = {10.1086/375026},
       adsurl = {https://ui.adsabs.harvard.edu/abs/2003ApJ...590.1111W}
}

\appendixpage
\begin{appendices}
\section{Vector sum, solar dipole and open flux}\label{appendix:VectorSum}
In the vector sum method, the pixels of a synoptic magnetogram are represented as vectors in heliographic spherical coordinates \citep{Tahtinen2024}.
The length of each pixel vector corresponds to the total signed magnetic flux within the magnetogram pixel and the direction to the location of the pixel on the solar surface.
The vector sum is straightforwardly the sum of these pixel vectors, which produces a vector sum, which we call here the open dipole flux (ODF) vector.
ODF vector shares its orientation with the first multipole (dipole) in the spherical harmonic expansion, but its magnitude differs from the solar dipole moment by a factor of $\frac{4\pi R_\odot^2}{3}$ and it has units of flux instead of flux density.

The magnitude of the open dipole flux vector closely matches the open solar flux (OSF) derived from the potential field source surface (PFSS) model \citep[see, e.g.,][ and references therein]{Wang1992}  with a source surface radius $R_{ss}=2.5R_\odot$.
\citet{Tahtinen2026a} also showed that the magnitude of the open dipole flux vector equals the photospheric magnetic flux aligned with the dipole axis, relating the PFSS OSF directly to the distribution of photospheric magnetic fields.

\end{appendices}

\end{document}